\documentclass[12pt]{article}

\usepackage{a4,amsmath,amssymb}

\newtheorem{theorem}{Theorem}

\title{On the NP-Completeness of Some Graph Cluster Measures} 

\author{
Ji\v{r}\'{\i} \v{S}\'{\i}ma
\thanks{Research partially supported by project 1M0021620808 of The
Ministry of Education of the Czech Republic.}\\
Institute of Computer Science,\\
Academy of Sciences of the Czech Republic,\\
P.\,O.~Box~5, 18207~Prague~8, Czech Republic,
\emph{sima@cs.cas.cz}
\and Satu Elisa Schaeffer
\thanks{Research supported by the Academy of Finland, grant 126235,
and the Nokia Foundation.}\\ 
Laboratory for Theoretical Computer Science,\\
Helsinki University of Technology, \\
P.~O.~Box 5400, FI-02015 TKK, Finland, 
\emph{elisa.schaeffer@tkk.fi}
}

\begin{document}

\maketitle

\begin{abstract}
Graph clustering is the problem of identifying sparsely connected
dense subgraphs (clusters) in a given graph. Proposed clustering
algorithms usually optimize various fitness functions that measure the
quality of a cluster within the graph. Examples of such cluster
measures include the conductance, the local and relative densities,
and single cluster editing. We prove that the decision problems
associated with the optimization tasks of finding the clusters that
are optimal with respect to these fitness measures are NP-complete. 
\end{abstract}

\section{Introduction}

Clustering is an important issue in the analysis and exploration of
data. There is a wide area of applications in data mining, VLSI
design, parallel computing, web searching, software engineering,
computer graphics, gene analysis, etc. See also~\cite{Jain99} for an
overview. Intuitively clustering consists in discovering natural
groups (clusters) of similar elements in data set. An important
variant of data clustering is graph clustering where the similarity
relation is expressed by a graph. In this paper, we restrict to
unweighted, undirected graphs with no self-loops.

We first recall some basic definitions from graph theory.
Let $G=(V,E)$ be an \emph{undirected} graph and denote by 
$E(S)=\{\{u,v\}\in E\,;\,u,v\in S\}$ the set of edges in a
\emph{subgraph} $G(S)=(S,E(S))$ \emph{induced} by a subset of
vertices $S\subseteq V$. We say that $S\subseteq V$ creates a
\emph{clique} of \emph{size} $|S|$ if edges in
$E(S)=\{\{u,v\}\,;\,u,v\in S,\,u\not=v\}$
join every two different vertices in $S$. Further denote by
$d_G(v)=|\{u\in V\,;$ $\{u,v\}\in E\}|$ the \emph{degree} of vertex
$v\in V$ in $G$. We say that graph $G$ is a \emph{cubic} graph if
$d_G(v)=3$ for every $v\in V$. Moreover, any subset of vertices
$A\subseteq V$ creates a \emph{cut} of $G$, that is a partition of $V$
into disjoint sets $A$ and $V\setminus A$. The \emph{size} of cut $A$
is defined as
\begin{equation}
c_G(A)=\left|\{\{u,v\}\in E\,;\,u\in A\,,v\in V\setminus A\}\right|\,,
\end{equation}
and 
\begin{equation}
d_G(S)=\sum_{v\in S}d_G(v)
\end{equation} 
denotes the sum of degrees in cut $S\subseteq V$. 
 
A canonical definition of a graph cluster does not exist, but it
is commonly agreed that a cluster should be a connected subgraph
induced by a vertex set $S$ with many internal edges $E(S)$ and few
edges to outside vertices in $V\setminus S$~\cite{Broder00,Kannan00}. 
In this paper we consider several locally computable fitness functions
that are used for measuring the quality of a cluster within the
graph. The prominent position among graph cluster measures is occupied
by the 
\emph{conductance}~\cite{Brandes03,Cheng05,Flake02,Gkantsidis03,Kannan00} 
which is defined for any cut $\emptyset\not=S\subseteq V$ in graph $G$
as follows
\begin{equation}
\label{conductS}
\Phi_G(S)=\frac{c_G(S)}{\min(d_G(S),d_G(V\setminus S))}\,.
\end{equation}
Furthermore, the \emph{local density} $\delta_G(S)$~\cite{Virtanen03} 
(cf.\ the \emph{average degree}~\cite{Holzapfel03}) of a subset
$\emptyset\not=S\subseteq V$ in graph $G$ is the ratio of the number
of edges in subgraph $G(S)$ induced by $S$ over the number of edges in
a clique of size $|S|$ vertices, that is
\begin{equation}
\label{ldensityS}
\delta_G(S)=\frac{|E(S)|}{\binom{|S|}{2}}
=\frac{2\cdot|E(S)|}{|S|\cdot(|S|-1)}
\end{equation}
for $S$ containing at least two vertices whereas define
$\delta_G(S)=0$ for $|S|=1$. Similarly, we define the \emph{relative
density}~\cite{Mihail02} of cut $\emptyset\not=S\subseteq V$ as
follows
\begin{equation}
\label{rdensityS}
\varrho_G(S)=\frac{|E(S)|}{|E(S)|+c_G(S)}\,.
\end{equation}
Yet another graph cluster measure which we call \emph{single cluster
editing} (cf.~\cite{Shamir02}) of a subset $S\subseteq V$ counts the
number of edge operations (both additions and deletions) needed to
transform $S$ into an isolated clique:
\begin{equation}
\label{scledit}
\varepsilon_G(S)=\binom{|S|}{2}-|E(S)|+c_G(S)\,.
\end{equation}

Proposed clustering
algorithms~\cite{Brandes03,Jain99,Mihail02,Schaeffer05} usually search 
for clusters that are optimal with respect to the above-mentioned
fitness measures. Therefore the underlying optimization problems of
finding the clusters that minimize the conductance or maximize the
densities or that need a small single cluster editing are of special
interest. In this paper we will formally prove that the associated
decision problems for the conductance (Section~\ref{conductance}),
local and relative densities (Section~\ref{densities}), and single
cluster editing (Section~\ref{editing}) are NP-complete. These
complexity results appear to be well-known or at least intuitively
credible, but not properly documented in the literature.

\section{Conductance}
\label{conductance}

Finding a subset of vertices that has the minimum conductance 
in a given graph has been often stated to be an NP-complete problem in
the literature 
\cite{Arora04,Brandes03,Carrasco03,Flake02,Kannan00,Leighton99,Lovasz96}. 
However, we could not find an explicit proof anywhere. For example,
the NP-completeness proof due to Papadimitrou~\cite{Shi00} for the
problem of finding the minimum \emph{normalized cut} which is in fact
the conductance of a weighted graph does not imply the hardness in the
unweighted case. Thus we provide the proof in this section. The
decision version for the conductance problem is formulated as follows: 

\medskip

\noindent
{\bf Minimum Conductance} ({\sc Conductance})\\
\emph{Instance:} An undirected graph $G=(V,E)$ and positive integer
$\phi$.\\  
\emph{Question:} Is there a cut $S\subseteq V$ such that
$\Phi_G(S)\leq\phi\,$?

\begin{theorem}
{\sc Conductance} is NP-complete.
\end{theorem}

\noindent\textbf{Proof:}
Clearly, {\sc Conductance} belongs to NP since a nondeterministic
algorithm can guess a cut $S\subseteq V$ and verify
$\Phi_G(S)\leq\phi$ in polynomial time. For the NP-hardness proof 
the following maximum cut problem on cubic graphs will be reduced to
{\sc Conductance} in polynomial time.

\medskip

\noindent
{\bf Maximum Cut for Cubic Graphs} ({\sc Max Cut--3})\\
\emph{Instance:} A cubic graph $G=(V,E)$ and positive integer $a$.\\ 
\emph{Question:} Is there a cut $A\subseteq V$ such that
$c_G(A)\geq a\,$?

\medskip

\noindent 
The {\sc Max Cut--3} problem was first stated to be NP-complete
in~\cite{Yannakakis78} which became a widely used
reference~\cite{Garey79} although an explicit proof cannot be found
there and we were unable to reconstruct the argument from the
sketch. Nevertheless, the NP-completeness of {\sc Max Cut--3}
follows from its APX-completeness presented in~\cite{Alimonti00}. The 
following reduction to {\sc Conductance} is adapted from that
used for the minimum edge expansion problem~\cite{Kaibel01}.

Given a {\sc Max Cut--3} instance, i.e.\  a cubic graph $G=(V,E)$
with $n=|V|$ vertices, and positive integer $a$, a corresponding
undirected graph $G'=(V',E')$ for {\sc Conductance} is composed of two
fully connected copies of the complement of $G$, that is 
$V'=V_1\cup V_2$ where $V_i=\{v^i\,;\,v\in V\}$ for $i=1,2$, and 
$E'=E_1\cup E_2\cup E_3$ where 
$E_i=\{\{u^i,v^i\}\,;\,u,v\in V,u\not=v,\{u,v\}\not\in E\}$ for
$i=1,2$, and $E_3=\{\{u^1,v^2\}\,;\,u,v\in V\}$. In addition, define
the required conductance bound
\begin{equation}
\label{phi}
\phi=\frac{1}{2n-4}\left(n-\frac{2a}{n}\right)\,.
\end{equation}
The number of vertices in $G'$ is $|V'|=2n$ and the number of edges
$|E'|=(2n-4)n$ since 
\begin{equation}
\label{degG'}
d_{G'}(v)=2n-4\quad\mbox{for every } v\in V'
\end{equation}
due to $G$ is a cubic graph. It follows that $G'$ can be constructed
in polynomial time.

For a cut $\emptyset\not=S\subseteq V'$ in $G'$ with $k=|S|\leq 2n$
vertices denote by 
\begin{equation}
\label{projS}
S_i=\{v\in V\,;\,v^i\in S\}\quad\mbox{for }i=1,2
\end{equation} 
the cuts in $G$ that are projections of $S$ to $V_1$ and $V_2$,
respectively. Since $c_{G'}(S)=c_{G'}(V'\setminus S)$ it holds 
$\Phi_{G'}(S)=\Phi_{G'}(V'\setminus S)$ according to definition
(\ref{conductS}). Hence, $k\leq n$ can be assumed without loss of
generality when computing the conductance in $G'$. Thus,
\begin{equation}
\Phi_{G'}(S)=\frac{|S|\cdot|V'\setminus S|-c_{G}(S_1)-c_{G}(S_2)}
{(2n-4)\cdot|S|}
\end{equation} 
follows from condition (\ref{degG'}) and the fact that $G'$ is
composed of two fully connected complements of $G$, which can be
rewritten as
\begin{equation}
\label{condG'2}
\Phi_{G'}(S)=
\frac{1}{2n-4}\left(2n-k-\frac{c_{G}(S_1)+c_{G}(S_2)}{k}\right)\,.
\end{equation} 

Now we verify the correctness of the reduction by proving that the 
{\sc Max Cut--3} instance has a solution if and only if the
corresponding {\sc Conductance} instance is solvable. First assume
that a cut $A\subseteq V$ exists in $G$ whose size satisfies 
\begin{equation}
\label{cuta}
c_{G}(A)\geq a\,. 
\end{equation}
Denote by
\begin{equation}
\label{SA}
S^A=\{v^1\in V_1\,;\, v\in A\}\cup
\{v^2\in V_2\,;\, v\in V\setminus A\}\subseteq V'
\end{equation} 
the cut in $G'$ whose projections (\ref{projS}) to $V_1$ and $V_2$ are
$S_1^A=A$ and $S_2^A=V\setminus A$, respectively. Since $|S^A|=n$ and  
$c_{G}(A)=c_{G}(V\setminus A)$ the conductance of $S^A$ can be upper
bounded as
\begin{equation}
\label{PhiSA}
\Phi_{G'}\left(S^A\right)=
\frac{1}{2n-4}\left(n-\frac{2c_{G}(A)}{n}\right)\leq 
\frac{1}{2n-4}\left(n-\frac{2a}{n}\right)=\phi
\end{equation} 
according to equations (\ref{condG'2}), (\ref{cuta}), and (\ref{phi}),
which shows that $S^A$ is a solution of the {\sc Conductance}
instance.

For the converse, assume that the conductance of cut
$\emptyset\not=S\subseteq V'$ in $G'$ meets
\begin{equation}
\label{condphi}
\Phi_{G'}(S)\leq\phi\,.
\end{equation}
Let $A\subseteq V$ be the maximum cut in $G$. For cut $S^A$ defined 
according to (\ref{SA}) we prove that
\begin{equation}
\label{PhiSAleqPhiS}
\Phi_{G'}\left(S^A\right)\leq\Phi_{G'}(S)
\end{equation}
which is rewritten to
\begin{equation}
\frac{1}{2n-4}\left(n-\frac{2c_{G}(A)}{n}\right)\leq
\frac{1}{2n-4}\left(2n-k-\frac{c_{G}(S_1)+c_{G}(S_2)}{k}\right)
\end{equation}
according to (\ref{PhiSA}) and (\ref{condG'2}) where $k=|S|\leq n$
and $S_1,S_2$ are defined in (\ref{projS}). Since
$2c_{G}(A)\geq c_{G}(S_1)+c_{G}(S_2)$ due to $A$ is the maximum cut in
$G$, it suffices to show
\begin{equation}
n-k+\left(\frac{1}{n}-\frac{1}{k}\right)(c_{G}(S_1)+c_{G}(S_2))\geq 0
\end{equation}
which follows from $\frac{1}{n}-\frac{1}{k}\leq 0$ and 
$c_{G}(S_1)+c_{G}(S_2)\leq |S_1|\cdot n+|S_2|\cdot n=kn$. Thus,
\begin{equation}
\frac{1}{2n-4}\left(n-\frac{2c_{G}(A)}{n}\right)=
\Phi_{G'}\left(S^A\right)\leq\Phi_{G'}(S)\leq\phi=
\frac{1}{2n-4}\left(n-\frac{2a}{n}\right)
\end{equation}
holds according to (\ref{PhiSA}), (\ref{PhiSAleqPhiS}),
(\ref{condphi}), and (\ref{phi}), which implies $c_G(A)\geq a$. Hence,
$A$ solves the {\sc MAX CUT-3} instance.~\hfill$\Box$

\section{Local and Relative Density}
\label{densities}

The decision version of the maximum density problem is formulated as
follows:

\medskip

\noindent
{\bf Maximum Density} ({\sc Density})\\
\emph{Instance:} An undirected graph $G=(V,E)$, positive integer
$k\leq |V|$, and a rational number $0\leq r\leq 1$.\\  
\emph{Question:} Is there a subset $S\subseteq V$ such that
$|S|=k$ and the density of $S$ in $G$ is at least $r\,$?

\medskip

\noindent
We distinguish between {\sc Local Density} and {\sc Relative Density}  
problems according to the particular density measure used which is
the local density (\ref{ldensityS}) and the relative density
(\ref{rdensityS}), respectively. Clearly, {\sc Local Density} is
NP-complete since this problem for $r=1$ coincides with the
NP-complete {\sc Clique} problem~\cite{Karp72}.
Also the NP-completeness of {\sc Relative Density} can easily be
achieved: 

\begin{theorem}
\label{reldensNP}
{\sc Relative Density} is NP-complete.
\end{theorem}

\noindent\textbf{Proof:}
Obviously, {\sc Relative Density} belongs to NP since a nondeterministic
algorithm can guess a cut $S\subseteq V$ of cardinality $|S|=k$ and
verify $\varrho_G(S)\geq r$ in polynomial time. For the NP-hardness proof 
the following minimum bisection problem on cubic graphs which is known
to be NP-complete~\cite{Bui87} will be reduced to {\sc Relative Density} in
polynomial time. 

\medskip

\noindent
{\bf Minimum Bisection for Cubic Graphs} ({\sc Min Bisection--3})\\
\emph{Instance:} A cubic graph $G=(V,E)$ with $n=|V|$ vertices and
positive integer $a$.\\  
\emph{Question:} Is there a cut $S\subseteq V$ such that $|S|=\frac{n}{2}$ 
and $c_G(S)\leq a\,$?

\medskip

\noindent 
Given a {\sc Min Bisection--3} instance, i.e.\  a cubic graph $G=(V,E)$
with $n=|V|$ vertices, and positive integer $a$, a corresponding 
{\sc Relative Density} instance consists of the same graph $G$,
parameters $k=\frac{n}{2}$ and 
\begin{equation}
\label{defr}
r=\frac{3n-2a}{3n+2a}\,.
\end{equation}
Now for any subset $S\subseteq V$ such that $|S|=k=\frac{n}{2}$
it holds
\begin{equation}
\label{cubES}
|E(S)|=\frac{3|S|-c_G(S)}{2}=\frac{3n-2c_G(S)}{4}
\end{equation}
due to $G$ is a cubic graph, which gives
\begin{equation}
\label{rhoGS}
\varrho_G(S)=\frac{3n-2c_G(S)}{3n+2c_G(S)}
\end{equation}
according to (\ref{rdensityS}). It follows from (\ref{defr}) and
(\ref{rhoGS}) that $\varrho_G(S)\geq r$ if\/f 
$c_G(S)\leq a$.~\hfill$\Box$

\section{Single Cluster Editing}
\label{editing}

The problem of deciding whether a given graph can be transformed into 
a collection of cliques using at most $m$ edge operations (both
additions and deletions) which is called {\sc Cluster Editing} is
known to be NP-complete~\cite{Shamir02}. When the desired solution
must contain exactly $p$ cliques, the so called 
{\sc p--Cluster Editing} problem remains NP-complete for every 
$p\geq 2$. Here we study the issue of whether a given graph contains a
subset $S$ of exactly $k$ vertices such that at most $m$ edge
additions and deletions suffice altogether to turn $S$ into an
isolated clique:

\medskip

\noindent
{\bf Minimum Single Cluster Editing} ({\sc 1--Cluster Editing})\\
\emph{Instance:} An undirected graph $G=(V,E)$, positive integers
$k\leq |V|$ and $m$.\\ 
\emph{Question:} Is there a subset $S\subseteq V$ such that
$|S|=k$ and $\varepsilon_G(S)\leq m\,$?

\medskip

\begin{theorem}
{\sc 1--Cluster Editing} is NP-complete.
\end{theorem}

\noindent\textbf{Proof:} Obviously, {\sc 1--Cluster Editing} belongs to
NP since a nondeterministic algorithm can guess a subset 
$S\subseteq V$ of cardinality $|S|=k$ and verify 
$\varepsilon_G(S)\leq m$ in polynomial time. For the NP-hardness proof
the {\sc Min Bisection--3} problem is used again (cf.\ the proof of
Theorem~\ref{reldensNP}) which will be reduced to 
{\sc 1--Cluster Editing} in polynomial time. 

Given a {\sc Min Bisection--3} instance, i.e.\  a cubic graph $G=(V,E)$
with $n=|V|$ vertices, and positive integer $a$, a corresponding 
{\sc 1--Cluster Editing} instance consists of the same graph $G$,
parameters $k=\frac{n}{2}$ and 
\begin{equation}
\label{defm}
m=\frac{12a+n(n-8)}{8}\,.
\end{equation}
Now for any subset $S\subseteq V$ such that $|S|=k=\frac{n}{2}$
it holds
\begin{equation}
\label{epsGS}
\varepsilon_G(S)=\frac{|S|\cdot(|S|-1)}{2}-\frac{3|S|-c_G(S)}{2}+c_G(S)
=\frac{12c_G(S)+n(n-8)}{8}
\end{equation}
according to (\ref{scledit}) and (\ref{cubES}). It follows from 
(\ref{defm}) and (\ref{epsGS}) that $\varepsilon_G(S)\leq m$ if\/f 
$c_G(S)\leq a$.~\hfill$\Box$

\section{Conclusion}

In this paper we have presented the explicit NP-completeness proofs 
for the decision problems associated with the optimization of four
possible graph cluster measures; namely the conductance,  
the local and relative densities, and single cluster editing. 
In clustering algorithms, combinations of fitness measures are often
preferred as only optimizing one may result in anomalies such as
selecting small cliques or connected components as clusters. An open
problem is the complexity of minimizing the {\em product} of the local
and relative densities~\cite{Schaeffer05} (e.g. their sum is closely
related to the edge operation count for the single cluster editing 
problem). Another important area for further research is the
complexity of finding related approximation solutions~\cite{Arora04}.

\end{document}